\documentclass[
 reprint,
 superscriptaddress,
 amsmath,
 amssymb,
 aps,
]{revtex4-2}

\usepackage{graphicx}
\usepackage{dcolumn}
\usepackage{bm}
\usepackage{physics}
\usepackage{float}
\usepackage{color}
\usepackage{comment}

\begin{document}

\preprint{APS/123-QED}

\title{Hardware-efficient quantum annealing with error mitigation via classical shadow}

\author{Takaharu Yoshida}
\email{mamesuke_117@icloud.com}
\thanks{T. Y. and Y. S. are equally contributed to this paper.}
\author{Yuta Shingu}
\email{shingu.yuta@gmail.com}

\author{Chihaya Shimada}

\author{Tetsuro Nikuni}
\affiliation{Department  of  Physics, Tokyo  University  of  Science,  Shinjuku,  Tokyo  162-8601,  Japan.}
\author{Hideaki Hakoshima}
\affiliation{Graduate School of Engineering Science, Osaka University, 1-3 Machikaneyama, Toyonaka, Osaka 560-8531, Japan}
\affiliation{Center for Quantum Information and Quantum Biology, Osaka University, 1-2 Machikaneyama, Toyonaka, Osaka 560-0043, Japan}
\author{Yuichiro Matsuzaki}
\affiliation{Department of Electrical, Electronic, and Communication Engineering, Faculty of Science and Engineering, Chuo University, 1-13-27 Kasuga, Bunkyo-ku, Tokyo, 112-8551, Japan}

\date{\today}

\begin{abstract}

Quantum annealing (QA) is an efficient method for finding the ground-state energy of the problem Hamiltonian. However, in practical implementation, the system suffers from decoherence. On the other hand, recently,  ``Localized virtual purification" (LVP) was
proposed to suppress decoherence in the context of noisy intermediate-scale quantum (NISQ) devices. 
Suppose observables have spatially local support in the lattice. In that case, the requirement for LVP is to calculate the expectation value with a reduced density matrix on a portion of the total system.
In this work, we propose a method to mitigate decoherence errors in QA using LVP. The key idea is to use the so-called classical shadow method to construct the reduced density matrix. Thanks to the CS, unlike the previous schemes to mitigate decoherence error for QA, we do not need either two-qubit gates or mid-circuit measurements, which means that our method is hardware-efficient.
\end{abstract}

\maketitle


\section{introduction}

Quantum annealing (QA)~\cite{kadowaki1998quantum,farhi2000quantum,farhi2001quantum,Ray1989-SK,finnila1994quantum,aspuru2005simulated} is a promising method for obtaining a ground state of a problem Hamiltonian $H_{\mathrm{p}}$. In QA, the ground state of a simple ``driver" Hamiltonian $H_{\mathrm{d}}$ is prepared, and a time-dependent Hamiltonian which changes from the driver Hamiltonian to the problem
Hamiltonian is employed. 
If an adiabatic condition is satisfied, there is a theoretical guarantee that the ground state of the problem Hamiltonian can be obtained from such a unitary dynamics~\cite{ehrenfest1916adiabatische,kato1950adiabatic,amin2009consistency,dodin2021generalized,Jansen2007-Bf}.

However, QA faces errors due to nonadiabatic transitions and decoherence effects~\cite{morita2008mathematical,albert1961quantum,albert1962quantum,roland2005noise,aaberg2005quantum,albash2015decoherence,childs2001robustness,sarandy2005adiabatic,ashhab2006decoherence}. When the dynamics is quick enough to suppress decoherence, it may result in
 unwanted transitions from the ground state to higher-energy excited states. Conversely, when the dynamics is slow enough to satisfy the adiabatic condition, decoherence becomes a dominant error factor, which can also cause unwanted transitions into excited states.
 This creates a trade-off between suppressing decoherence and avoiding nonadiabatic transitions, reducing the accuracy of the expectation value.

On the other hand, a significant amount of theoretical and experimental effort has been devoted to achieving practical applications with noisy intermediate-scale quantum (NISQ) computing devices~\cite{preskill2018quantum,endo2021hybrid,cerezo2021variational}. NISQ devices enable quantum computations with hundreds to potentially thousands of qubits, with gate error rates of around $10^{-3}$ or lower~\cite{barends2014superconducting,brown2011single}. Numerous algorithms have been developed specifically for NISQ computing. In many research studies, these algorithms utilize variational quantum circuits to create a trial wave function that minimizes a cost function~\cite{peruzzo2014variational,li2017efficient,mcclean2016theory,yuan2019theory,endo2020variational}. To accomplish this, measurements of observables on the qubits, aligned with the trial wave function, are required. However, due to noise within real-world devices, obtaining precise expectation values for these observables is challenging, degrading performance. Fortunately, sophisticated techniques known as ``quantum error mitigation” (QEM) reduce the effect of environmental noise~\cite{endo2021hybrid,li2017efficient,kandala2019error,temme2017error,endo2018practical,song2019quantum,ou2020characterization,mcardle2019error,bonet2018low,sun2021mitigating,larose2022mitiq,google2020hartree,mcclean2017hybrid,stark2014self,greenbaum2015introduction,blume2017demonstration,strikis2021learning,czarnik2021error,wang2021scalable,o2021error,yoshioka2022generalized,cai2022quantum,cao2022mitigating,huo2021self,takagi2021optimal,hakoshima2021relationship,hakoshima2024localized,suzuki2022quantum,li2024ensemble}. These techniques incorporate extra quantum gates and post-processing, which mitigate the impact of noise and improve computational accuracy. Since QA also suﬀers from noise-induced errors, applying QEM to QA has been an important research direction. The previous work~\cite{shingu2024quantum} proposed a QEM-based approach to mitigate decoherence eﬀects in QA, demonstrating its theoretical feasibility.

 The virtual purification method, also known as the exponential error suppression method, can reduce errors without detailed knowledge of the noise model~\cite{koczor2021exponential,huggins2021virtual,czarnik2021qubit,yamamoto2021error}. In this paper, we call this method "full-size" virtual purification (FVP). FVP utilizes the $n$ copies of the noisy quantum state $\rho$ to mitigate errors and approximate the ideal quantum circuit. Assuming that each noisy state consists of $N$ qubits, and these copies are produced by the same quantum circuit and subject to the same noise model, we apply entangling gates across the copies and obtain $\ev{O}^{(n)}_{\mathrm{FVP}}=\mathrm{Tr}[O\rho^{(n)}_{\mathrm{FVP}}]$ for an observable $O$, where $\rho^{(n)}_\mathrm{FVP} = \rho^n /\mathrm{Tr}[\rho^n]$. A key benefit of this method is that the population of the dominant state in $\rho$  approximates unity, which means that the error is suppressed exponentially as we increase the number of copies $n$. However, the FVP method requires $Nn$ qubits and controlled gates, and the huge measurement cost is also needed. These features make it challenging to implement the FVP method into the current quantum devices.

Recently, an alternative method called “localized virtual purification" (LVP)  was proposed to alleviate this difficulty~\cite{hakoshima2024localized}.
Especially this method is helpful to estimate the ground-state energy of the target (problem) Hamiltonian if the interactions in the Hamiltonian are geometrically local.
LVP is a method of virtual purification on local subsets of qubits and we can ignore the regime of the system, which is far from the support of observable $O$ to be measured. 
In general, the output of LVP is not equivalent to that of FVP, but these difference is exponentially small assuming the clustering property, which can be derived from the Lieb-Robinson bound~\cite{lieb1972finite,hastings2004lieb,nachtergaele2006lieb,hastings2010locality}, known as the upper bounds of a velocity of the information propagation. 
We denote the density matrix reduced to the local area as $\rho_{\mathrm{LVP}}$, and then LVP leads to
$\operatorname{Tr}[\rho_{\mathrm{LVP}}O]\approx \operatorname{Tr}[\rho_{\mathrm{ideal}}O]$, where $\rho_{\mathrm{ideal}}=|\psi _{\mathrm{ideal}}\rangle \langle \psi_{\mathrm{ideal}}|$ denotes a state without decoherence.
The required number of two-qubit gates between copies
and measurement shots for LVP are much smaller than that for FVP, because FVP applies the entangling gates among copies over the extensive portion of the total qubits while FVP cares about the far local area.

In this paper, we propose a method of QA with LVP mitigate the decoherence effect. The key idea is to replace the SWAP gates used in the original LVP scheme with classical shadow (CS) measurements. CS was proposed to estimate observables with fewer measurements and used to efficiently measure the purity of the state~\cite{huang2020predicting,paini2021estimating,chen2021robust}. Using CS instead of entangling gates, we can obtain the reduced density matrix and calculate the expectation values of individual terms of local Hamiltonian and purity, which will be placed into the numerator and the denominator respectively for LVP. Since LVP exploits the locality of Hamiltonian, this
approach significantly reduces computational cost compared with the FVP with CS~\cite{seif2023shadow}, which requires exponential computational cost.
Furthermore, unlike the previous QEM schemes for QA ~\cite{shingu2024quantum}, 
our method does not require projective measurements in the middle of the process but only at the end, making it compatible with the current hardware. To verify the effectiveness of our scheme, we performed numerical calculations using the spin-1/2 XXZ model as a problem Hamiltonian, and demonstrated that our method provides more accurate the ground-state energy estimates.
This method is also applicable to the wider range of analogue quantum simulators to mitigate the effects of decoherence, which are inevitable for all real-world quantum devices.
\begin{figure*}
    \centering
    \includegraphics[width=\linewidth]{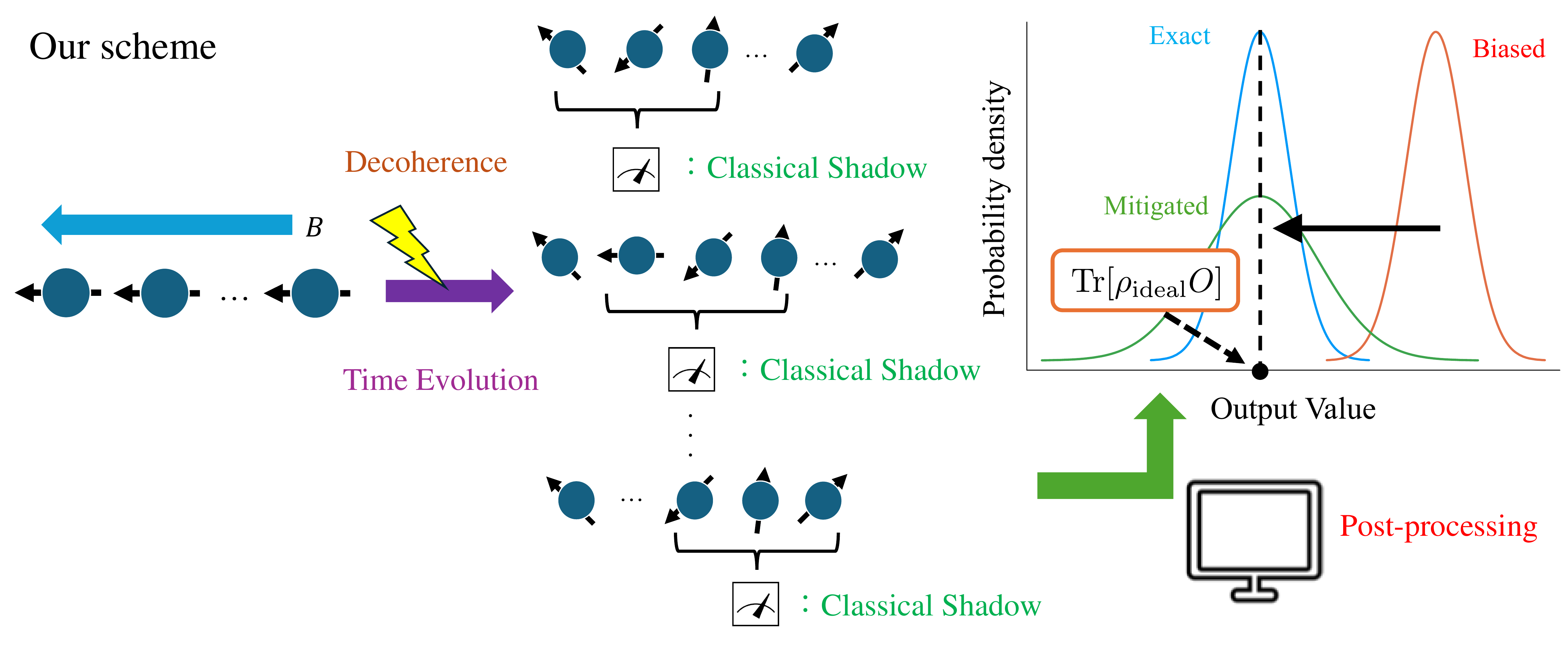}
\caption{Schematic diagram of our schemes in QA. We first perform QA on generally noisy devices, similar to conventional QA methods, and then apply classical-shadow tomography to obtain reduced density matrices. Afterward, we move on to the post-processing phase, where we substitute these density matrices into Eq.(\ref{LVP-QA}). The symbol $O$ represents an observables of interest in and $\rho_{\mathrm{ideal}}$ is the ideal density matrix.}
    \label{fig:EMQACS-method}
\end{figure*}

\section{Quantum annealing and the effect of decoherence}
 In this section, we review QA~\cite{kadowaki1998quantum,farhi2000quantum,farhi2001quantum,Ray1989-SK,finnila1994quantum} for obtaining the ground state and the ground-state energy of a problem Hamiltonian. The total Hamiltonian of an annealing-based quantum device is expressed as
 \begin{equation}
     H(t)=A(t)H_\mathrm{p} + B(t)H_\mathrm{d} ,\label{qahamiltonian}
 \end{equation}
 where $H_\mathrm{d}$ is the driver Hamiltonian
and $H_\mathrm{p}$ is the problem Hamiltonian.
 Additionally, $A(t)$ and $B(t)$ are time-dependent coefficients. The coefficients of the total QA Hamiltonian are given by
 \begin{equation}
 \begin{split}
    A(t)&=\frac{t}{T},\\
    B(t)&=1-\frac{t}{T},
 \end{split}
 \end{equation}
 where $T$ is the annealing time. 
The initial state is chosen to be the ground state of the driver Hamiltonian.

Under the dynamics governed by the Hamiltonian in Eq.~\eqref{qahamiltonian}, the adiabatic theorem~\cite{ehrenfest1916adiabatische,kato1950adiabatic,amin2009consistency,dodin2021generalized, Jansen2007-Bf} guarantees that one can obtain the ground state of the problem Hamiltonian provided that the adiabatic condition is satisfied.

The density matrix following QA is given by  $\rho = \sum_j p_j |E_j\rangle\langle E_j|$where $p_j$ represents the population and $|E_j\rangle$ denotes an eigenvector of the problem Hamiltonian with the energy eigenvalue $E_j$. When the problem is of the Ising type, consisting solely of $\sigma_z$, the eigenvector $\ket{E_j}$ corresponds to the computational basis. Therefore, measuring in the computational basis after QA allows the ground state to be obtained with the probability given by the ground state population. If the ground-state population is nonzero, increasing the number of trials enables us to obtain the ground state in a practical computation time. In contrast, when the problem Hamiltonian includes off-diagonal matrix elements, its energy cannot be estimated through measurements in the computational basis alone. In such cases, Pauli measurements are required. Specifically, we consider a Hamiltonian expressed as $H = \sum_i c_i \sigma_i$, where $\sigma_i$ represents a Pauli product and $c_i$ is its corresponding coefficient. The expectation value of each term is obtained by measuring the Pauli products in the quantum states after QA, and the total energy is determined by summing these expectation values. For a density matrix $\rho$, the energy expectation value is given by $H = \text{Tr} [H\rho] = \sum_n p_n E_n$, which is obtained by performing many measurements. If there is a nonzero population in the excited states, the experimentally measured energy will deviate from the true ground-state energy. Therefore, to accurately determine the ground-state energy, it is essential to prepare a density matrix that closely approximates the exact ground state.

In the ideal case, QA is a promising method for estimating the ground-state energy. However, there are two main challenges to its implementation in practice: environmental decoherence and nonadiabatic transition. While performing QA for a long time can mitigate the effect of nonadiabatic transitions, it also makes quantum states more susceptible to environmental decoherence. To account for the effect of decoherence, we use the Gorini–Kossakowski–Sudarshan–Lindblad (GKSL)  master equation, which describes the dynamics of the system~\cite{lindblad1976generators,gorini1976completely}:
\begin{equation}
    \frac{d \rho}{d t}=-i[H(t), \rho]+\sum_k \frac{\lambda_k}{2}\left(2L_k\rho L_k^{\dagger}-\left\{L_k^{\dagger} L_k, \rho\right\}\right),
\end{equation}
where $[\bullet,\bullet]$ denotes the commutator, $\lambda_n$ denotes a decay rates, $L_n$ denotes the Lindblad operator, and $\{\bullet,\bullet\}$ denotes the anticommutator. 

There is a wide range of research aimed at minimizing nonadiabatic transitions and decoherence during QA. One approach involves using an inhomogeneous driver Hamiltonian to accelerate QA for specific problem Hamiltonians~\cite{susa2018exponential,susa2018quantum}. On the other hand, Seki et al. demonstrated that the efficiency of QA can be enhanced for certain types of problem Hamiltonians by employing ``nonstochastic" Hamiltonian with negative off-diagonal matrix elements~\cite{seki2012quantum,seki2015quantum}. It has also been understood that the energy gap between the ground and first excited states can be robustly estimated to mitigate nonadiabatic transitions~\cite{matsuzaki2021direct,russo2021evaluating,mori2024experimentally}. Additionally, various strategies have been developed to reduce environmental noise. One such method uses error correction with ancillary qubits to reduce decoherence during QA~\cite{pudenz2014error}, while another involves a decoherence-free subspace~\cite{suzuki2020proposal}. Spin-lock techniques are also advantageous for maintaining long-lived qubits in QA~\cite{chen2011experimental,nakahara2013lectures,matsuzaki2020quantum}. Furthermore, several approaches have been explored to enhance QA performance by harnessing nonadiabatic transitions, quenching~\cite{crosson2014different,goto2020quantum,hormozi2017nonstoquastic,muthukrishnan2016tunneling,brady2017necessary,somma2012quantum,das2008colloquium,karanikolas2020pulsed}, and degenerating two-level systems~\cite{watabe2020enhancing}. Variational methods have also been used to reduce nonadiabatic transitions and decoherence in QA~\cite{susa2021variational,matsuura2021variationally,passarelli2021transitionless,imoto2022quantum}. 
Recently, several QA methods incorporating QEM have been proposed~\cite{sun2021mitigating,amin2023quantum,schiffer2024virtual,shingu2024quantum}. One of these studies mainly focused on the coherent nonadiabatic transitions, and demonstrated the effectiveness of the method. 
The other papers explored
methods to mitigate decoherence. 


\section{Localized Virtual Purification}
In this section, we review a QEM method called localized virtual purification (LVP)~\cite{hakoshima2024localized}. First, we introduce the FVP method from which LVP originated. On realistic quantum devices, obtained state is a mixed state described as $\rho$, which can be expand by the complete orthonormal bases $\{\ket{\Psi_i}\}^{2^N-1}_{i=0}$ as follows
\begin{equation}
\label{afterQA}
    \rho = p\ket{\Psi_0}\bra{\Psi_0} + (1-p)\sum_{k=1}^{2^N-1}{c_k\ket{\Psi_k}\bra{\Psi_k}}.
\end{equation}
Here, $N$ is the number of qubits, and $p$ is the population of $\ket{\Psi_0}$, which satisfies $p>0$. We  assume $c_k>0$ and $\sum_kc_k=1$. The state $\ket{\Psi_0}$ represents the eigenstate of $\rho$ with the largest eigenvalue, i.e., $p>(1-p)c_k$ for all $k$. 
In this paper, we assume that our goal is to prepare the ground state $\ket{E_0}$ of the problem Hamiltonian, and $\ket{\Psi_0}$ a good approximation of this state, i.e.,
$|\bra{\Psi_0}\ket{E_0}|\approx 1$ is satisfied.
The expectation value of an observable $\hat{A}$ for mixed state in Eq. (\ref{afterQA}) becomes
\begin{equation}
    \mathrm{Tr}[\hat{A}\rho] = p\bra{\Psi_0}\hat{A}\ket{\Psi_0} + (1-p)\sum_{k=1}^{2^N-1}c_k\bra{\Psi_k}\hat{A}\ket{\Psi_k},
\end{equation}
which deviates from the true expectation value $\langle E_0|\hat{A}|E_0\rangle $ due to decoherence.
To suppress the effect of decoherence, virtual purification methods virtually construct a purified density matrix with $n$ copies of the state $\rho$:
\begin{equation}
\label{FVP:def}
\rho^{(n)}_{\mathrm{FVP}}=\frac{\rho^n}{\mathrm{Tr}[\rho^n]}.
\end{equation}
The expectation value of an observable $\hat{A}$ then becomes
\begin{eqnarray}
\label{FVP-expect}
    \operatorname{Tr}[\rho^{(n)}_{\mathrm{FVP}}\hat{A}]&=& \frac{p^n}{\mathrm{Tr}[\rho^n]}\bra{\Psi_0}\hat{A}\ket{\Psi_0} \nonumber \\
    &+& \frac{(1-p)^n}{\mathrm{Tr}[\rho^n]}\sum_{k=1}^{2^N-1}c_k^{n}\bra{\Psi_k}\hat{A}\ket{\Psi_k}.
\end{eqnarray}
Since $p^n/\mathrm{Tr}[\rho^n]>p$ for $p>1/2$, this methods can effectively suppresses error. Moreover, since $p^n/\mathrm{Tr}[\rho^n] \rightarrow 1 \ (n\rightarrow\infty)$ in Eq.~(\ref{FVP-expect}), the decoherence effects are exponentially suppressed by increasing $n$ if its strength is sufficiently low. However, implementing FVP requires controlled gates on all qubits across multiple copies, which is expensive for NISQ devices. Additionally, calculating the denominator  in Eq. (\ref{FVP:def}) given by $1/\operatorname{Tr}[\rho^n]$ incurs a high sampling cost due to the global support of $\rho$, making its estimation challenging. 

LVP is a recently proposed method to circumvent these difficulties. In the following discussion, we consider the Hamiltonian with local interactions. Instead of $\ev{H_{\mathrm{p}}}$, we consider the following expectation value to estimate the ground-state energy:
\begin{equation}
\left\langle H_{\mathrm{p}}\right\rangle_{\mathrm{LVP}}^{(n)}=\sum_i \frac{\operatorname{Tr}_{A_i+B_i}\left[\rho_{A_i+B_i}^n H_i\right]}{\operatorname{Tr}_{A_i+B_i}\left[\rho_{A_i+B_i}^n\right]},
\label{LVP-QA}
\end{equation}
where $H_i$ is the local term of the problem Hamiltonian defined as $H_{\mathrm{p}}=\sum_iH_i$. Here, $A_i$ denotes the support of $H_i$ and $B_i$ denotes the region surrounding $A_i$. Also, $\rho_{A_i+B_i}\equiv\operatorname{Tr}_{C_i}[\rho]$, where $C_i$ represents the complement set of $A_i\cup B_i$ and $\operatorname{Tr}_{C_i}[\bullet]$ denotes the operation of tracing out the region $C_i$. The deviation of the expectation value of LVP from that of FVP can be written as follows:
\begin{equation}
D^{(n)}_i\left(H_i\right)=\frac{\operatorname{Tr}_{A_i+B_i}\left[\rho_{A_i+B_i}^n H_i\right]}{\operatorname{Tr}_{A_i+B_i}\left[\rho_{A_i+B_i}^n\right]}-\frac{\operatorname{Tr}\left[\rho^n H_i\right]}{\operatorname{Tr}\left[\rho^n\right]}.\label{Deviation}
\end{equation}

Let us denote the total deviation from FVP as $|D^{(n)}\left(H_{\mathrm{p}}\right)|$, which is bounded by the sum of each term $|D^{(n)}\left(H_{\mathrm{p}}\right)|\le\sum_i|D^{(n)}_i\left(H_i\right)|$. Using the fact that Eq.(\ref{Deviation}) can be rewritten in the form of the generalized correlation function, and assuming that the $H_\mathrm{p}$ is gapped, meaning that the energy difference between the ground-state and first excited-state energies is finite, one can prove that the following inequality holds:
\begin{equation}
\label{LVPtheo1}
\left|D_0^{(n)}\left(H_{\mathrm{p}}\right)\right| \leq c\sum_i\left\|H_{i}\right\| \frac{\left\|\rho_{C_i}^{n-1}\right\|}{\operatorname{Tr}\left[\rho_{C_i}^n\right]} \exp \left(-\frac{d(A_i, C_i)}{\xi_i}\right),
\end{equation}
where $\xi_i$ and $c$ are constants independent of the system size $N$, and $d(A_i,C_i)$ represents the distance between the regions $A_i$ and $C_i$, which defined as the minimum distance among the qubits in $A_i$ and $C_i$.This inequality suggests that LVP is effective when Hp is gapped, as the energy gap leads to exponential suppression of deviations from FVP. Specifically, in systems with local interactions and sufficiently large distances between regions $A_i$ and $C_i$, LVP mitigates decoherence effects, making it a promising method for such systems.

\section{Localized virtual purification in quantum annealing}

In this section, we describe our QA scheme for estimating the ground-state energy of a local problem Hamiltonian while mitigating environmental noise effects, as the schematic diagram shown in Fig \ref{fig:EMQACS-method}. To adapt the QEM method discussed in the previous section to QA, we employ the CS method~\cite{huang2020predicting,paini2021estimating,chen2021robust} to calculate Eq.~(\ref{LVP-QA}), as implementing 2-qubit gates is challenging for current QA devices. Throughout this paper, we assume that the Pauli-basis measurements are used for CS. If we apply the FVP method to the QA with the CS, the numerous number of measurement shots in the order of $\exp(O(N))$ is needed. In contrast, the required number of measuremen
shots for LVP is much smaller as we will see below.

In this paper, we focus the one-dimensional system. We start from the driver Hamiltonian, and consider the resulting noisy state described by the density matrix 
in Eq.(\ref{afterQA}). To implement the LVP method, we need to calculate the reduced density matrix $\rho_{A_i+B_i}$ for each $H_i$. For this purpose, we apply the CS method to the regions $A_i$ and $B_i$, and we obtain $\rho^n_{A_i+B_i}/\operatorname{Tr}[\rho^n_{A_i+B_i}]$ after classical post-processing. Since the Eq.~(\ref{LVPtheo1}) is satisfied, the deviation between the expectation value with LVP and that with FVP is exponentially small.
Here, we aim to bound the deviation to an arbitrary small $\varepsilon$. To achieve this, we need to set $d(A_i,C_i)$ to be $O(\log N)$
order. In this case, $|D^{(n)}(H_{\mathrm{p}})|$ is well-bounded because $\|H_{\mathrm{p}}\|\|\rho_C^n\|/\operatorname{Tr}[\rho_c^n]$ grows at most with $O(N)$ order in one-dimensional system, while $\exp(-d(A_i,C_i)/\xi)$ decays with $O(\operatorname{poly}(N))$. After QA, we carry out CS to obtain $\rho_{A_i+B_i}$, and since the number of qubits in the region $B_i$ denoted as $N_B$ also scales logarithmically with $N$, we can compute both the numerator $\operatorname{Tr}[\rho_{A_i+B_i}^2H_i]$ and the denominator $\operatorname{Tr}[\rho_{A_i+B_i}^2]$ of Eq.(\ref{LVP-QA}) in polynomial order, assuming a constant number of copies, such as $n=2$. To compute the purity $\operatorname{Tr}[\rho_{A_i+B_i}^2]$, additional sampling cost is required when using CS instead of two-qubit gates. However, this cost is at most polynomial because we only care about the limited fraction of the total system. Therefore, the total cost of CS remains $O(\operatorname{poly}N)$. While we have discussed only about one-dimensional system, our scheme may also work well practically in two or higher-dimensional systems as the original paper~\cite{hakoshima2024localized} proposing LVP method suggests.

Note that our scheme is more feasible for QA devices compared with the previous studies that applied an QEM scheme to QA. Some previous studies~\cite{schiffer2024virtual,shingu2024quantum} require 
the so-called mid-circuit measurements, which involve performing projective measurements in the middle of calculation. However, this requirement is challenging for certain quantum computing platforms. For the quantum annealing device provided by D-wave, the superconducting qubit can be measured using superconducting quantum interference device (SQUID)~\cite{clarke2006squid}, but the quantum state will be destroyed after the measurements. For neutral atoms~\cite{saffman2016quantum,barredo2016atom}, conventional fluorescence measurement destroys the quantum state, like with superconducing qubits. Current QuEra devices are not capable of implementing the mid-circuit measurement, and even in the near future, this will remain expensive because mid-circuit measurements require additional laser and impose long-time readout~\cite{graham2023midcircuit}. An alternative methods~\cite{sun2021mitigating,amin2023quantum}, which do not require the mid-circuit measurement, need the capability of energy rescaling. However, energy rescaling poses challenges for the scalability of Hamiltonian simulations, particularly in neutral atom simulators, making it difficult to implement on these devices. These methods are also challenging for other types of quantum devices. On the other hand, our method does not require mid-circuit measurements or energy-rescaling because we only perform Pauli measurements for CS after normal QA dynamics.

\section{results}
In this section, we present numerical results to evaluate the performance of our proposed method. For this purpose, As the problem Hamiltonian in QA, we consider the one-dimensional spin-1/2 XXZ Hamiltonian with nearest-neighbor interactions and uniform magnetic field along the $y$-direction, given by
\begin{equation}
    H_{\mathrm{p}}=J\sum_{i=1}^N(\sigma^x_i\sigma^x_{i+1} +\sigma^y_i\sigma^y_{i+1} + \Delta\sigma^z_i\sigma^z_{i+1}) + h\sum_{i=1}^N\sigma_i^y,
\end{equation}
with periodic boundary conditions where $J$ denotes the coupling strength, $h$ denotes the amplitude of the magnetic field, and $\Delta$ denotes the isotropy parameter. In out numerical simulations, we set $J=-1$, $h=1$, and $\Delta=-0.73$. As the driver Hamiltonian, we adopt the random transverse field Hamiltonian, given by $H_\mathrm{d} = -\sum_i{h_i\sigma_i^x}$ where $h_i$ is the amplitude of the magnetic fields and random variables following a uniform distribution $\mathrm{Uniform}[0,1]$, and we start QA from the ground-state of $H_{\mathrm{d}}$. To incorporate the effect of noise, we describe the system time-evolution using the GKSL master equation with depolarizing noise:
\begin{equation}
\frac{d \rho}{d t}=-i[H(t), \rho]-\frac{\lambda}{2} \sum_{i=1}^N \sum_{j \in\{x, y, z\}}\left[\hat{\sigma}_i^j,\left[\hat{\sigma}_i^j, \rho\right]\right],
\label{GKSL-depo}
\end{equation}
where we set $\lambda=0.0025$ in numerical simulations.
As the metric to assess the performance of our scheme, we define the relative error as $(\ev{H_{\mathrm{p}}}-E_g)/|E_g|$ where $\ev{H_{\mathrm{p}}}$ is the expectation value of the problem Hamiltonian obtained from the numerical simulations, and $E_g$ denotes the ground-state energy of the problem Hamiltonian obtained by exact diagonalization. We solved Eq.~(\ref{GKSL-depo}) using qutip library in Python, and obtained the density matrix. After that, we calculated the reduced density matrix by taking partial trace over the $C_i$, which is the complement set of $A_i\cup B_i$. Here, $A_i$ is the support of $H_i$, which is defined as
\begin{equation}
    H_i=J(\sigma^x_i\sigma^x_{i+1} +\sigma^y_i\sigma^y_{i+1} + \Delta\sigma^z_i\sigma^z_{i+1}) + h\sigma_i^y,
\end{equation}
and $B_i$ is the regime bordering on $A_i$ whose area is decided by $d(A_i,C_i)$.

\begin{figure}[t]
    \centering
    \includegraphics[width=\linewidth]{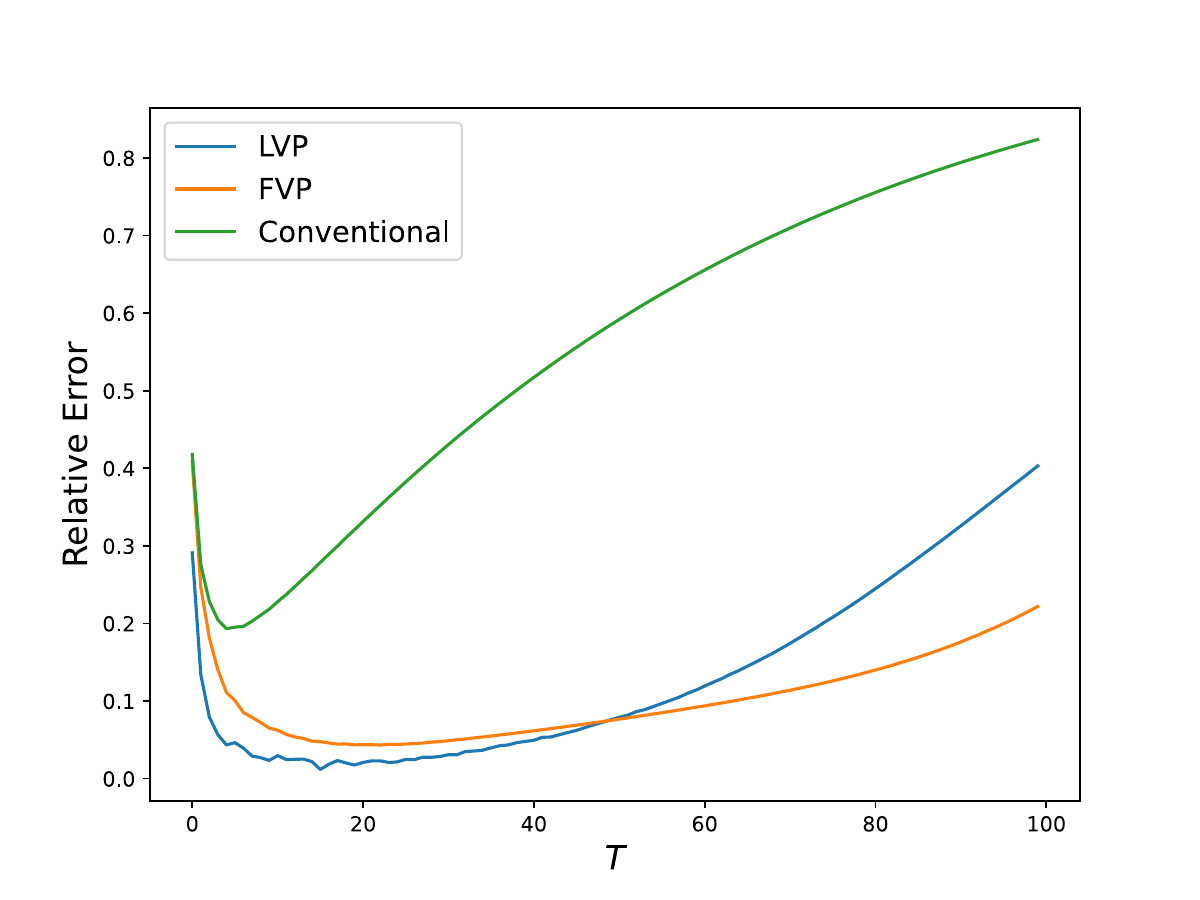}
    \caption{Relative error of the ground-state energy of the problem Hamiltonian $H_{\mathrm{p}}$ with $N=9$.  The relative error is defined as $(\langle H_{\mathrm{p}}\rangle -E_g)/|E_g|$ where $\langle H_{\mathrm{p}}\rangle$ denotes the expectation value with the state obtained from each QA method. and $E_g$ denotes the actual ground-state energy obtained from exact diagonalization.}
    \label{fig:Rel_error}
\end{figure}

In Fig.~\ref{fig:Rel_error}, we show the numerical result of  the relative error as a function of the annealing time $T$ comparing our LVP-based method with FVP and QA without QEM with system size $N=9$. In Fig.~\ref{fig:Rel_error}, we ignored the process of tomography to evaluate the performance in ideal case, which the calculation considering about the tomography process is expected to approach asymptotically by increasing the number of shots. The results reveal that there is an optimal value of $T$ where the expectation values are minimized, rather than asymptotically approaching the true ground-state energy for large $T$. This behavior emerges because decoherence effects become more pronounced as $T$ increases, while nonadiabatic transitions are suppressed. For the LVP method, we take $n=2$ copies and set $d(A_i,C_i)=1$, which corresponds $N_{B_i}=2$ in one-dimensional systems. Our results confirm that the LVP-based method suppresses the effect of errors as effectively as FVP. This is the significant advantage of our method, considering the fact that LVP with CS does not require any two-qubit gates unlike the FVP. On the other hand, we observed that the minimum expectation value of LVP can be lower than the exact ground-state energy, while the energy calculated with $\operatorname{Tr}[H_\mathrm{p}\rho]$ always satisfies $\operatorname{Tr}[\rho H_\mathrm{p}] \ge E_\mathrm{g}$ for an arbitrary density matrix $\rho$. This is because the ``expectation value" obtained in the LVP-based method is defined in terms of locally purified version of the reduced density matrix, as given in Eq.~(\ref{LVP-QA}), which is not necessarily expressible in the same form $\operatorname{Tr}[H_\mathrm{p} \rho]$. Consequently, minimizing the expectation value of the energy by changing $T$ does not necessarily yield the closest approximation to the true value. However, as discussed earlier, the deviation from the exact energy can be arbitrary smaller by increasing $d(A_i,C_i)$, which is increased by increasing the number of qubits to measure.

Before closing this section, we highlight the potential applicability of our method beyond QA. While this paper has focused on QA implementations, our method is applicable for broader class of analogue quantum simulators, provided that the problem Hamiltonian is local, meaning that it contains only the terms with a local support. Therefore, our proposal has broader utility than demonstrated in this work. 

\begin{table}[b]
    \centering
    \caption{The minimum expectation value $\langle{H_\mathrm{p}}\rangle$ obtained by each algorithm with $\lambda$ = 0.005, along with the exact value of the ground-state energy.}
    \label{tab:energies}
    \begin{tabular}{|c|c|c|c||c|}
        \hline
         & LVP & FVP & Conventionl & Exact\\ \hline
        $N=6$ & -14.144 & -13.413 & -11.212 & -14.058\\ \hline
        $N=7$ & -16.533 & -16.072 & -14.410 & -16.388\\ \hline
        $N=8$ & -18.431 & -17.762 & -14.179 & -18.729\\ \hline
        $N=9$ & -20.820  & -20.155 & -16.996 & -21.068\\ \hline
    \end{tabular}
\end{table}

\section{summary and future directions}
In this paper, we have proposed a QEM technique for QA
based on the LVP method and the CS with low measurement costs. Our method improves upon the FVP and original LVP approaches by eliminating the need for two-qubit gates making it more suitable for current QA devices. We numerically demonstrated that using our protocol allows us to estimate the ground-state energy with high accuracy compared with conventional QA under decoherence.

The future direction include extending LVP to long-ranged Hamiltonians, where interactions decay with distance according polynomial inverse power law, such as van der Waals interaction $r^{-6}$~\cite{browaeys2016experimental}, considering the extension of Lieb-Robinson bounds to long-ranged models~\cite{matsuzaki2021direct}. While extension will impose more severe costs on measurement and limitation on the applicable class of Hamiltonian, it is still reasonable to expect that upper bound of information spreading may help us to reduce the overall cost.

\section*{Acknowledgement}
This work is supported by MEXT Quantum Leap Flagship Program (MEXT Q-LEAP) Grant No. JPMXS0120319794.
This work was supported by JST Moonshot R\&D (Grant Number JPMJMS226C).
YM is supported by JSPS KAKENHI (Grant Number 23H04390), JST Presto
(Grant No. JPMJPR245B) Japan, and CREST(Grant No. JPMJCR23I5).

\bibliographystyle{apsrev4-2}
\bibliography{EMQAwCSref}

\end{document}